\documentclass[final,11pt]{article}
\usepackage{multirow}
\usepackage[nottoc]{tocbibind}
\usepackage{geometry}
\usepackage{subcaption}
\usepackage[affil-it, auth-sc]{authblk}
\usepackage[english]{babel}
\usepackage[utf8]{inputenc}
\usepackage{cite}
\usepackage{graphicx}
\usepackage{amssymb}
\usepackage{amsthm}
\usepackage{amsmath}
\usepackage{amsfonts}
\usepackage{verbatim}
\usepackage{lineno}
\usepackage{float}
\usepackage{todonotes}
\presetkeys{todonotes}{inline, color=blue!30, size=\small}{}
\usepackage{color, soul}
\definecolor{darkred}{rgb}{0.9, 0.0, 0.0}
\definecolor{darkgreen}{rgb}{0.0, 0.5, 0.0}
\usepackage[colorlinks,bookmarks,linkcolor=darkred, citecolor=darkgreen]{hyperref}
\usepackage{eso-pic}
\usepackage[sort&compress,numbers]{natbib}
\usepackage{doi}
\usepackage{paralist,epsfig}
\usepackage{relsize}
\usepackage{nicefrac,esint}
\usepackage{float}
\usepackage{cleveref}
\usepackage{feynmp-auto}
\usepackage{slashed}

\newcommand{\hp}{{\frac{1}{2}}}
\newcommand{\hm}{{-\frac{1}{2}}}

\def\slash#1{#1\!\!\!/\!\,\,}

\def\fAt{f_{A3}}
\def\fAtbar{\overline{f}_{A3}}

\begin{document}

\AddToShipoutPictureFG*{
 \AtPageUpperLeft{\put(-60,-60){\makebox[\paperwidth][r]{LA-UR-23-33998, FERMILAB-PUB-24-0534-T, CETUP2025-012}}}}
 
\title{Resonance Contributions to Radiative Corrections in Charged-Current Elastic (Anti)Neutrino-Nucleon Scattering at GeV Energies}

\author{Oleksandr Tomalak\thanks{tomalak@itp.ac.cn}}
\affil{Theoretical Division, Los Alamos National Laboratory, Los Alamos, NM 87545, USA}
\affil{Institute of Theoretical Physics, Chinese Academy of Sciences, Beijing 100190, China}

\date{\today}

\maketitle
We present the first evaluation of virtual resonance contributions to the charged-current elastic (anti)neutrino-nucleon scattering at GeV energies, focusing on the dominant $\Delta(1232)$ resonance. We approximate the vector part of the $N \to \Delta$ transition by the leading magnetic dipole term. Our results for the cross-section corrections at fixed neutrino energy indicate the permille-level contribution of resonance intermediate states to the elastic and radiative scattering cross sections. This calculation exhibits the expected infrared behavior of the invariant amplitudes and unpolarized cross sections. Our findings provide important insights into inelastic excitations in the charged-current elastic (anti)neutrino-nucleon scattering at GeV energies.
\tableofcontents

\section{Introduction}
\label{sec:Introduction}

Unpolarized electron elastic scattering is a well-established tool for studying electromagnetic properties of nucleons and nuclei~\cite{Hofstadter:1953zjy,Hofstadter:1955ae,Rosenbluth:1950yq,A1:2010nsl,A1:2013fsc,Punjabi:2015bba,Mihovilovic:2019jiz,Xiong:2019umf,Christy:2021snt,A1:2021njh,A1:2022wzx,Qattan:2024pco}. Modern measurements of electron-proton elastic scattering at GeV energies and below are performed with precision at the percent level or better. Such accuracy calls for a careful treatment of radiative corrections~\cite{McKinley:1948zz,Mo:1968cg,Barbieri:1972as,Barbieri:1972hn,Vanderhaeghen:2000ws,Maximon:2000hm,Bonciani:2003te,Arbuzov:2015vba,Hill:2016gdf,Arbuzov:2019hcg,Afanasev:2021nhy,Engel:2023arz} and nucleon excitations. In particular, the diagram with two exchanged photons is a subject of active theoretical and experimental investigation~\cite{Gunion:1972bj,Carlson:2007sp,Arrington:2011dn,Afanasev:2023gev}. It has attracted increasing attention~\cite{Gunion:1972bj,Blunden:2003sp,Guichon:2003qm} after the first extraction of the proton electromagnetic form-factor ratio from polarization transfer data~\cite{Akhiezer:1968ek,Akhiezer:1973xbf,Dombey:1969wk,Dombey:1969wi,JeffersonLabHallA:1999epl,JeffersonLabHallA:2001qqe,Punjabi:2005wq,Puckett:2010ac,Puckett:2017flj,GEp-IIIGEp-2Gamma:2017ftw,SANE:2018cub}, which resulted in contradiction with a traditional Rosenbluth separation technique at GeV electron beam energies. At these energies, on top of effective field theory considerations~\cite{Hill:2011wy,Hill:2012rh,Dye:2016uep,Dye:2018rgg,Talukdar:2019dko,Talukdar:2020aui,Cao:2021nhm,Peset:2021iul,Choudhary:2023rsz}, there are two well-developed ways to approach the two-photon exchange diagram: hadronic models and dispersion relations. Calculations within the hadronic model~\cite{Blunden:2003sp,Blunden:2005ew,Kondratyuk:2005kk,Kondratyuk:2007hc,Hill:2011wy,Graczyk:2013pca,Lorenz:2014yda,Zhou:2014xka,Tomalak:2014dja} assume the on-shell form for the photon-proton interaction and perform a straightforward evaluation of the loop diagram. Such an evaluation is ultraviolet-finite and collinear-safe, but it violates unitarity for the real parts of the two-photon exchange amplitudes. This violation is not significant for the proton, or elastic, intermediate state, when the lepton mass is neglected~\cite{Tomalak:2014sva,Tomalak:2018jak}, but it leads to cross sections that diverge with beam energy when modeling the resonance intermediate states~\cite{Tomalak:2016vbf,Tomalak:2017shs}. The dispersion-relation approach~\cite{Gorchtein:2006mq,Borisyuk:2008es,Borisyuk:2012he,Borisyuk:2013hja,Tomalak:2014sva,Borisyuk:2015xma,Tomalak:2018jak}, based on the on-shell information only, does not suffer from such pathological behavior and is considered a physically-motivated and favorable framework for the evaluation of the two-photon exchange diagram in electron-proton scattering community~\cite{Blunden:2017nby}. As of today, proton and pion-nucleon intermediate states in the two-photon exchange diagram are included in the dispersion-relation approach. Such a calculation takes experimentally measured proton form factors~\cite{A1:2010nsl,A1:2013fsc} and invariant amplitudes of pion electroproduction~\cite{Drechsel:1998hk,Drechsel:2007if} as an input. Accounting for the pion-nucleon intermediate states improves agreement between theoretical predictions and recent experimental data from electron- and positron-proton scattering experiments~\cite{Rachek:2014fam,CLAS:2013mza,CLAS:2014xso,CLAS:2016fvy,OLYMPUS:2016gso,OLYMPUS:2020dgl,Schmidt:2019vpr}. The remaining discrepancy might be explained by the resonant enhancement of nucleon excitations at GeV beam energies~\cite{Afanasev:2017gsk}, when the propagator of the resonance intermediate state moves close to its pole position.

A similar diagram with the exchange of two bosons contributes to the elastic (anti)neutrino-nucleon scattering. To fully benefit from experimental advances in the fields of neutrino scattering and oscillation experiments~\cite{DUNE:2021tad,MINERvA:2023avz,Petti:2023abz},
this diagram was recently revisited by a few independent groups~\cite{DeRujula:1979grv,Day:2012gb,Graczyk:2013fha,Tomalak:2021hec,Tomalak:2022xup}. As a result, it was included in a new and complete factorization framework for radiative corrections to the charged-current elastic (anti)neutrino-nucleon scattering in Refs.~\cite{Tomalak:2021hec,Tomalak:2022xup}. The latter references evaluated the nucleon intermediate state and provided a model consistent with the expected soft and collinear behavior. The authors estimated the uncertainty from inelastic intermediate states as a contribution from the diagrams with electromagnetic coupling to the neutron. As in electron scattering, resonance intermediate states in the charged-current (anti)neutrino elastic scattering might be enhanced for the incoming beam energies that move the resonance propagators close to their pole positions. That is why resonance contributions may be relevant for precision analyses of (anti)neutrino scattering data.

In this paper, we perform the first evaluation of the $\Delta (1232)$-resonance contribution to the unpolarized charged-current elastic (anti)neutrino-nucleon scattering at GeV energies. We verify that fixed-$t$ dispersion relations do not apply to resonance contributions as they are for the elastic intermediate state~\cite{Tomalak:2021hec,Tomalak:2022xup}. Therefore, we calculate the loop integrals by modeling the $\Delta$ interaction vertex in its on-shell form with two choices: (1) shifting momenta to satisfy collinear constraints as in the model of Refs.~\cite{Tomalak:2021hec,Tomalak:2022xup} and (2) retaining the full momentum dependence in the loop integration as in the hadronic model of Refs.~\cite{Blunden:2003sp,Blunden:2005ew,Kondratyuk:2005kk,Kondratyuk:2007hc,Graczyk:2013pca,Lorenz:2014yda,Zhou:2014xka,Tomalak:2014dja}, cf. also an interesting discussion in Ref.~\cite{Plestid:2025ojt}. Our model satisfies soft and collinear constraints on the real parts of the invariant amplitudes. It also respects the expected crossing properties for several contributions, specified in Section~\ref{sec:virtual_delta}, and fulfills the unitarity constraints for the imaginary parts. Additionally, we investigate soft and collinear properties of the invariant amplitudes and study off-shell effects by changing the vector part of the weak interaction vertex of the $\Delta$ resonance from the gauge-independent form of the magnetic dipole contribution only to a traditional in neutrino physics form-factor decomposition.

The paper is organized as follows. In Section~\ref{sec:delta_production}, we describe the production of the $\Delta$ resonance in electroweak processes at tree level and specify the amplitudes for the $\Delta$ neutrinoproduction. We provide the invariant amplitude decomposition and general cross-section expression in the elastic (anti)neutrino-nucleon scattering in Section~\ref{sec:elastic_scattering}. In Section~\ref{sec:virtual_delta}, we evaluate the virtual contribution of the $\Delta$ intermediate state to the elastic (anti)neutrino-nucleon scattering amplitudes and cross sections and provide numerical results for incoming (anti)neutrino beam energies around GeV scale. In Section~\ref{sec:factorization_framework}, we discuss $\Delta$ intermediate state contribution to radiative corrections in the factorization framework and corresponding radiation of real photons. Conclusions and outlook are presented in Section~\ref{sec:conclusions}. In Appendix~\ref{app:helicity_and_invariants}, we specify relations between invariant and helicity amplitudes in the elastic (anti)neutrino-nucleon scattering.

\section{$\Delta$ production in charged-current (anti)neutrino-nucleon scattering}
\label{sec:delta_production}

In this Section, we describe the details of the $\Delta$-resonance production in the charged-current (anti)neutrino-nucleon scattering at tree level.

We consider all possible channels for the production of the $\Delta$ resonance in the charged-current scattering of the neutrino $\nu_\ell$ and antineutrino $\overline{\nu}_\ell$ on the neutron $n$ and proton $p$: 1) $\nu_\ell n \to \ell^- \Delta^+$; 2) $\nu_\ell p \to \ell^- \Delta^{++}$; 3) $\overline{\nu}_\ell n \to \ell^+ \Delta^-$; 4) $\overline{\nu}_\ell p \to \ell^+ \Delta^0$. We denote the momentum of the incoming (anti)neutrino as $p$, the momentum of the outgoing charged lepton as $p^\prime$, the momentum of the initial-state nucleon as $k$, and the momentum of the final-state resonance as $k^\prime$. The kinematics of the resonance production can be completely described by the invariant mass of the resonance $W$, where $W^2 = \left( k^\prime \right)^2$, and two Mandelstam variables: the squared momentum transfer $Q^2 = - \left( k^\prime - k \right)^2$ and the squared energy in the (anti)neutrino-nucleon center-of-mass reference frame $s=\left( k + p \right)^2$.

At leading order in quantum electrodynamics (QED), the matrix element for the charged-current neutrinoproduction of the $\Delta$ resonance is expressed in terms of the $N \to \Delta$ transition current $J^{\mu \nu}_{N \to \Delta}$ as
\begin{align}
	T_{\nu_\ell N \to \ell^- \Delta} &= \sqrt{2} G_F V_{ud} \overline{\ell}^- \gamma_\mu \mathrm{P}_\mathrm{L} \nu_{\ell} \overline{\Psi}_\nu J^{\mu \nu}_{N \to \Delta} N, \label{eq:delta_neutrinoproduction1} \\
	T_{\overline{\nu}_\ell N \to \ell^+ \Delta} &= \sqrt{2} G_F V^*_{ud} \overline{\overline{\nu}}_\ell \gamma_\mu \mathrm{P}_\mathrm{L} \ell^+ \overline{\Psi}_\nu J^{\mu \nu}_{N \to \Delta} N, \label{eq:delta_neutrinoproduction2}
\end{align}
with the Fermi coupling constant $G_F$~\cite{Fermi:1934hr,Feynman:1958ty,vanRitbergen:1999fi,MuLan:2012sih}, the matrix element of the Cabibbo-Kobayashi-Maskawa (CKM) quark mixing matrix $V_{ud}$~\cite{Cabibbo:1963yz,Kobayashi:1973fv,Hardy:2020qwl,ParticleDataGroup:2020ssz}, and the Rarita-Schwinger spinor $\Psi_\nu \left( p_\Delta, \lambda_\Delta \right)$~\cite{Rarita:1941mf}. The $N \to \Delta$ transition current $J^{\mu \nu}_{N \to \Delta}$ can be expressed as a sum of vector $J^{\mu \nu}_{V, N \to \Delta}$ and axial-vector $J^{\mu \nu}_{A, N \to \Delta}$ contributions $J^{\mu \nu}_{N \to \Delta} = i \left( J^{\mu \nu}_{V, N \to \Delta} \gamma_5 + J^{\mu \nu}_{A, N \to \Delta} \right)$ with a Lorentz-invariant decomposition involving the transition form factors $C_3^{V,A},~C_4^{V,A},~C_5^{V,A}$, and $C_6^{V,A}$~\cite{Albright:1964sgs,Albright:1965aud,Salin:1967as,Bijtebier:1970ku,Zucker:1971hp,LlewellynSmith:1971uhs,Schreiner:1973ka,Schreiner:1973mj,Sato:2003rq,Lalakulich:2006sw}:\footnote{Our normalization reproduces Figure 9 of Ref.~\cite{Lalakulich:2005cs}.}
\begin{align}
	- i J^{\mu \nu}_{V, N \to \Delta} \gamma_5 &= \frac{g^{\nu \mu} \slash{q} -q^\nu \gamma^\mu}{M} C_3^V + \frac{g^{\nu \mu} q \cdot \left( p + q \right) -q^\nu \left( p^\mu + q^\mu \right)}{M^2} C_4^V + \frac{ g^{\nu \mu} q \cdot p -q^\nu \left( p^\mu + q^\mu \right)}{M^2} C_5^V + g^{\nu \mu} C_6^V, \label{eq:form_factors_decomposition1} \\
	J^{\mu \nu}_{A, N \to \Delta} &= \frac{g^{\nu \mu} \slash{q} -q^\nu \gamma^\mu}{M} C_3^A+ \frac{g^{\nu \mu} q \cdot \left( p + q \right) -q^\nu \left( p^\mu + q^\mu \right)}{M^2} C_4^A + g^{\nu \mu} C_5^A + \frac{q^{\nu} q^{\mu} }{M^2} C_6^A. \label{eq:form_factors_decomposition2}
\end{align}

For the vector contribution, we assume the conservation of the vector current and restrict ourselves to the leading magnetic dipole transition
\begin{equation}
	J^{\mu \nu}_{V, N \to \Delta} = \sqrt{\frac{3}{2}} \frac{\left( M+ W \right) {G}^*_\mathrm{M} \left( Q^2 \right) }{M \left(\left( M+ W \right)^2 + Q^2 \right)} \varepsilon^{\nu \mu \rho \sigma} \left( k^\prime \right)_\rho {q}_\sigma, \qquad q = k^\prime - k,\label{eq:vector_transition_form_factors}
\end{equation}
with the on-shell magnetic transition form factor $ G^*_\mathrm{M} \left( Q^2 \right) $, in the Jones-Scadron convention~\cite{Jones:1972ky}. For the form factor $G^*_\mathrm{M} \left( Q^2 \right)$, we take a large-$N_c$ theory relation~\cite{Pascalutsa:2006up,Pascalutsa:2007wz} in terms of the proton and neutron elastic Pauli form factors $ F_2^p$ and $ F_2^n$, respectively:
\begin{align}
	G^*_\mathrm{M} \left( Q^2 \right) & = \frac{G^*_\mathrm{M} \left( 0 \right) }{\mu_p - \mu_n - 1} \left[ F^p_2 (Q^2) - F^n_2 (Q^2) \right], \qquad G^*_\mathrm{M} \left( 0 \right) = 3.02, \\
	F^p_2 (Q^2) & = \frac{\mu_p - 1}{\left( 1 + \tau \right) \left( 1 + \frac{Q^2}{\Lambda^2} \right)^2}, \qquad \mu_p = 2.793, \qquad \Lambda = 0.843~\mathrm{GeV}, \qquad \tau = \frac{Q^2}{4 M^2}, \\
	F^n_2 (Q^2) & = \frac{\mu_n}{\left( 1 + \tau \right) \left( 1 + \frac{Q^2}{\Lambda^2} \right)^2} \frac{ 1 + \left( a + b \right) \tau}{ 1 + b \tau } , \quad \mu_n = -1.913, \quad a = 1.25, \quad b = 18.3, \label{eq:delta_GM}
\end{align}
where the neutron electric form factor is taken from Ref.~\cite{JeffersonLabE93-026:2003tty}. For the neutron magnetic, proton electric, and proton magnetic form factors, a dipole form is assumed. Using the magnetic dipole transition approximation of Eq.~(\ref{eq:vector_transition_form_factors}), the coupling to virtual photons becomes gauge-invariant and the unpolarized $\Delta$ production cross section does not contain vector-axial-vector interference terms. In this approximation, the vector transition form factors in Eq.~(\ref{eq:form_factors_decomposition1}) are given by
\begin{align}
	C^V_3 \left( Q^2 \right) &= \sqrt{\frac{3}{2}} \frac{W \left( M+ W \right)}{ \left( M+ W \right)^2 + Q^2 } {G}^*_\mathrm{M} \left( Q^2 \right), \label{eq:three_vector_transition_form_factors1} \\
	C^V_4 \left( Q^2 \right) &= - \sqrt{\frac{3}{2}} \frac{M \left( M+ W \right)}{ \left( M+ W \right)^2 + Q^2 } \frac{\sqrt{ \left( W - M \right)^2 + Q^2}}{\sqrt{ \left( W + M \right)^2 + Q^2}} {G}^*_\mathrm{M} \left( Q^2 \right), \label{eq:three_vector_transition_form_factors2} \\
	C^V_5 \left( Q^2 \right) &= \sqrt{\frac{3}{2}} \frac{W \left( M+ W \right)}{ \left( M+ W \right)^2 + Q^2 } \left( 1 - \frac{\sqrt{ \left( W - M \right)^2 + Q^2}}{\sqrt{ \left( W + M \right)^2 + Q^2}} \right) {G}^*_\mathrm{M} \left( Q^2 \right). \label{eq:three_vector_transition_form_factors3}
\end{align}
The conservation of the vector current implies $C^V_6 = 0$.

In Refs.~\cite{Adler:1968tw,Lalakulich:2005cs,Hernandez:2007qq}, the common assumptions for the axial-vector transition form factors consist of the following $3$ equations:
\begin{equation}
	C^A_3 \left( Q^2 \right) = 0 , \qquad
	C^A_4 \left( Q^2 \right) = - \frac{C^A_5 \left( Q^2 \right)}{4} , \qquad
	C^A_6 \left( Q^2 \right) = \frac{M^2}{m^2_\pi + Q^2} C^A_5 \left( Q^2 \right), \label{eq:delta_model}
\end{equation}
which we call the $\Delta_\mathrm{I}$ model. We also consider an alternative approximation $C^A_4 \left( Q^2 \right)= 0$ of Ref.~\cite{Fogli:1979cz}, which we call the $\Delta_\mathrm{II}$ model. For numerical estimates in this paper, we take the axial-vector transition form factor $C^A_5 \left( Q^2 \right)$ from Ref.~\cite{Lalakulich:2005cs}:
\begin{equation} \label{eq:CA5_modelling}
	C^A_5 \left( Q^2 \right) = \frac{C_5^A \left( 0 \right)}{\left( 1 + \frac{Q^2}{M^2_A} \right)^2} \frac{1}{ 1+ 2 \frac{Q^2}{M^2_A}},
\end{equation}
with the fixed $C_5^A \left( 0 \right) = 1.2$, in agreement with the large-$N_c$ analysis of Ref.~\cite{Karl:1984cz}, and a fit parameter $M^2_A = 1.05~\mathrm{GeV}^2$. According to Ref.~\cite{Lalakulich:2005cs}, such a form factor provides a better agreement with experimental data. Moreover, the functional form in Eq.~(\ref{eq:CA5_modelling}) satisfies the perturbative QCD scaling of the form factor $C_5^A \left( Q^2 \right) \lesssim \left( \ln Q \right)/Q^6$, which we obtain by generalizing the derivations in Refs.~\cite{Lepage:1980fj,Chernyak:1983ej,Carlson:1985mm}.\footnote{We also obtain the perturbative QCD scaling for the other axial-vector transition form factors: $C_3^A \left( Q^2 \right) \lesssim \left( \ln Q \right)/Q^8$ and $C_4^A \left( Q^2 \right) \lesssim \left( \ln Q \right)/Q^8$.}

\section{Charged-current elastic (anti)neutrino-nucleon scattering}
\label{sec:elastic_scattering}

In this Section, we describe the invariant amplitude decomposition in the elastic (anti)neutrino-nucleon scattering for the charged-current processes.

The matrix elements of the charged-current elastic processes $T^{m_\ell = 0}_{\nu_\ell n \to \ell^- p}$ and $T^{m_\ell = 0}_{\overline{\nu}_\ell p \to \ell^+ n}$ with a massless charged lepton can be expressed as~\cite{Tomalak:2021hec,Tomalak:2022xup}
\begin{align}
	T^{m_\ell = 0}_{\nu_\ell n \to \ell^- p} &= \sqrt{2} G_F V_{ud} \overline{\ell}^- \gamma^\mu \mathrm{P}_\mathrm{L} \nu_{\ell}\, \overline{p} \left(\gamma_\mu \left( {g}_M + {f}_A \gamma_5 \right) - \left( f_2 - 2 \fAt \gamma_5 \right) \frac{{K}_\mu}{M} \right) n, \label{eq:CCQE_amplitude1} \\
	T^{m_\ell = 0}_{\overline{\nu}_\ell p \to \ell^+ n} &= \sqrt{2} G_F V^*_{ud} \overline{\overline{\nu}}_\ell \gamma^\mu \mathrm{P}_\mathrm{L} \ell^+
\overline{n} \left( \gamma_\mu \left( \overline{g}_M + \overline{f}_A \gamma_5 \right) - \left( \overline{f}_2 + 2 \fAtbar \gamma_5 \right) \frac{{K}_\mu}{M} \right) p, \label{eq:CCQE_amplitude2}
\end{align}
with the averaged nucleon momentum $K_\mu = (k_\mu + k^\prime_\mu)/2$. Accounting for the nonzero lepton mass, four more invariant amplitudes are required to describe the matrix elements of the charged-current elastic processes, i.e., $T_{\nu_\ell n \to \ell^- p} = T^{m_\ell = 0}_{\nu_\ell n \to \ell^- p} + T^{m_\ell \neq 0}_{\nu_\ell n \to \ell^- p}$ and $T_{\overline{\nu}_\ell p \to \ell^+ n} = T^{m_\ell = 0}_{\overline{\nu}_\ell p \to \ell^+ n} + T^{m_\ell \neq 0}_{\overline{\nu}_\ell p \to \ell^+ n}$~\cite{Tomalak:2024yvq,Borah:2024hvo}:
\begin{align}
	T^{m_\ell \neq 0}_{\nu_\ell n \to \ell^- p} &= \sqrt{2} G_F V_{ud} \frac{m_\ell}{M} \left[ \frac{{f}_{T}}{4} \overline{\ell}^- \sigma^{\mu \nu} \mathrm{P}_\mathrm{L} \nu_{\ell}\, \overline{p} \sigma_{\mu \nu} n - \overline{\ell}^- \mathrm{P}_\mathrm{L} \nu_{\ell}\, \overline{p} \left( {f}_3 + f_P \gamma_5 - \frac{f_R}{4} \frac{\slash{P}}{M} \gamma_5 \right) n \right], \label{eq:CCQE_amplitudem1} \\
	T^{m_\ell \neq 0}_{\overline{\nu}_\ell p \to \ell^+ n} &= \sqrt{2} G_F V^\star_{ud} \frac{m_\ell}{M} \left[ \frac{\overline{f}_{T}}{4} \overline{\overline{\nu}}_\ell \sigma^{\mu \nu} \mathrm{P}_\mathrm{R} \ell^+ \overline{n} \sigma_{\mu \nu} p - \overline{\overline{\nu}}_\ell \mathrm{P}_\mathrm{R} \ell^+
\overline{n} \left( \overline{f}_3 + \overline{f}_P \gamma_5 - \frac{\overline{f}_R}{4} \frac{\slash{P}}{M} \gamma_5 \right) p\right], \label{eq:CCQE_amplitudem2}
\end{align}
with the averaged lepton momentum $P_\mu = (p_\mu + p^\prime_\mu)/2$. The invariant amplitudes are functions of two kinematic variables: the crossing-symmetric variable $\nu = E_\nu / M - \tau - r_\ell^2$, with $r_\ell = m_\ell/(2M)$, and the momentum transfer $Q^2 = - \left( p - p^\prime \right)^2 = - \left( k^\prime - k \right)^2$.

The corresponding general expression for the charged-current elastic scattering cross section (without radiation) can be written as~\cite{LlewellynSmith:1971uhs}\footnote{We neglect the relative difference in nucleon masses, $(m_n-m_p)/(m_n+m_p)$, and electroweak power corrections suppressed by the $W$-boson mass, $Q^2/M_W^2$; these effects contribute at the permille level.}
\begin{equation}
	\frac{d\sigma}{dQ^2} (E_\nu, Q^2) = \frac{G_F^2 |V_{ud}|^2}{2\pi} \frac{M^2}{E_\nu^2} \left[ \left( \tau + r_\ell^2 \right)A(\nu,~Q^2) - \frac{\nu}{M^2} B(\nu,~Q^2) + \frac{\nu^2}{M^4} \frac{C(\nu,~Q^2)}{1+ \tau} \right] ,\label{eq:xsection_CCQE}
\end{equation}
with the structure-dependent quantities $A$, $B$, and $C$~\cite{Tomalak:2024yvq,Borah:2024hvo}:
\begin{align}
	A &= \tau | g_M |^2 - | g_E |^2 + (1+ \tau) | f_A |^2 - r_\ell^2 \left( | g_M |^2 + | f_A + 2 f_P |^2 - 4 \left( 1 + \tau \right) \left( |f_P|^2 + |f_3|^2\right)\right) - 4 \tau (1+ \tau) |\fAt|^2 \nonumber \\
& - 2 r_\ell^2 \mathfrak{Re} \Big[ \left( g_E + 2 g_M - 2 \left( 1 + \tau \right) \fAt \right) f_T^\star \Big] - \eta r_\ell^2 \left( 1 + \tau + r_\ell^2 \right) \mathfrak{Re}\Big [ f_A f_R^\star \Big] - r_\ell^2 \left( 1 + 2 r_\ell^2 \right) |f_T|^2 \nonumber \\
&- 2 \eta r_\ell^4 \mathfrak{Re}\Big [ f_P f_R^\star \Big]+ \frac{r_\ell^2}{4} \left( 1 + \tau + \nu^2 - \left(1 + \tau + r_\ell^2 \right)^2 \right) |f_R|^2, \\
	B &= \mathfrak{Re} \Big[ 4 \eta \tau f^*_A g_M - 4 \eta r_\ell^2 \left( f_A - 2 \tau f_P \right)^* \fAt + 4 r_\ell^2 g_E f_3^\star - 2 \eta r_\ell^2 \left( 3 f_A - 2 \tau \left( f_P - \eta f_3\right) \right) f_T^\star \Big] \nonumber \\
&- r_\ell^4 \mathfrak{Re} \Big[ \left( f_T + 2 \fAt \right) f_R^\star \Big], \\
	C &= \tau |g_M|^2 + |g_E|^2 + (1 + \tau) |f_A|^2 + 4 \tau (1 + \tau) |\fAt|^2 + 2 r_\ell^2 \left( 1 + \tau \right) |f_T|^2 + \eta r_\ell^2 \left( 1 + \tau \right) \mathfrak{Re} \Big[ f_A f_R^\star \Big],
\end{align}
with $\eta = + 1$ for neutrino scattering and $\eta = - 1$ for antineutrino scattering.

Following the tree-level notation for the form factors, the electric and magnetic amplitudes $g_E$ and $g_M$ are defined from $f_1$ and $f_2$ as
\begin{equation}
	g_E = f_1 - \tau f_2, \qquad g_M = f_1 + f_2.
\end{equation}

At tree level, four amplitudes can be expressed in terms of the isovector vector, axial-vector, and induced pseudoscalar nucleon form factors $g_M = G^V_M,~g_E = G^V_E,~f_A=F_A,$ and $f_P = F_P$, while the other four amplitudes vanish. We will use a standard ansatz (partially conserved axial-vector current and the assumption of the pion pole dominance) for the pseudoscalar form factor: $F_P(Q^2) = 2M^2 F_A(Q^2)/\left(m_\pi^2 + Q^2\right)$.

\section{$\Delta$ intermediate state in charged-current elastic (anti)neutrino-nucleon scattering}
\label{sec:virtual_delta}

In this Section, we evaluate the contribution of virtual QED diagrams with the $\Delta$ intermediate state to invariant amplitudes and unpolarized cross sections by performing model calculations of one-loop direct and crossed box integrals.\footnote{As for the nucleon intermediate state contribution, the high-energy behavior of imaginary parts of the contribution from the spin-$3/2$ resonance to the charged-current elastic process does not allow us to write down fixed-$t$ unsubtracted dispersion relations for any of the invariant amplitudes.}

\begin{figure}[t]
\centering
\includegraphics[width=0.2\textwidth]{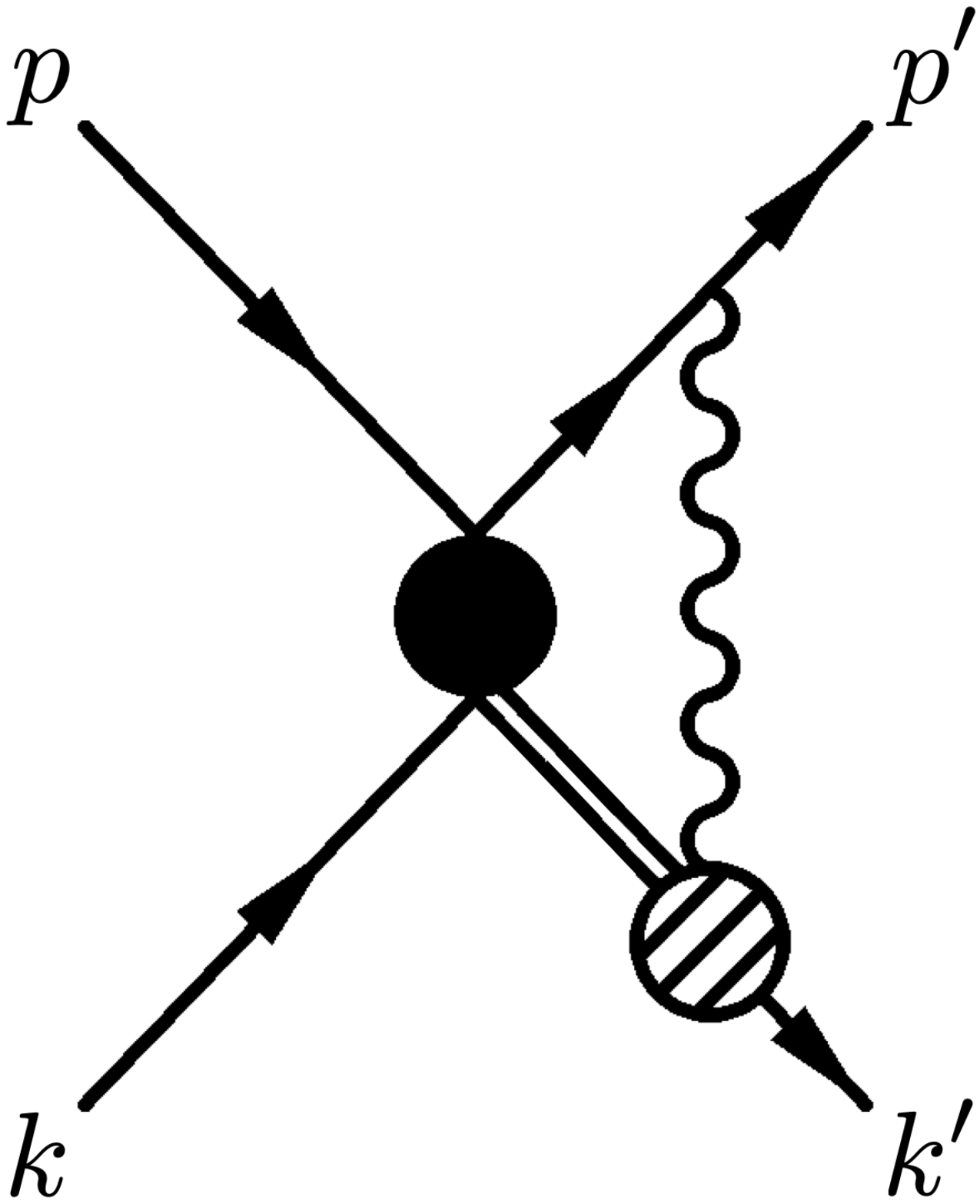}
\qquad \qquad
\includegraphics[width=0.2\textwidth]{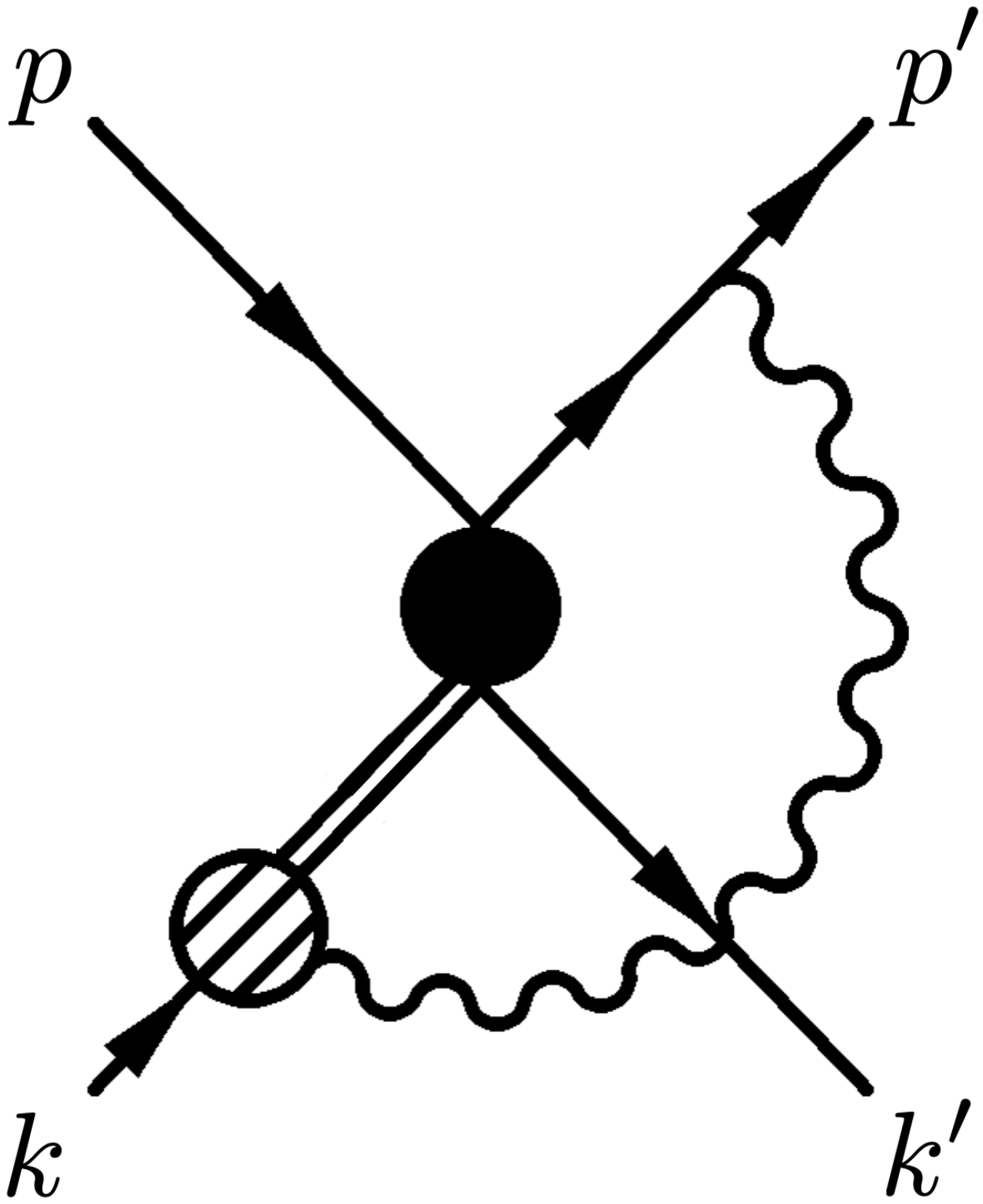}
\caption{Contributions of the $\Delta$ resonance to the charged-current elastic (anti)neutrino-nucleon scattering are shown. The photon is exchanged between the charged lepton and $N \to \Delta$ transition vertex. \label{fig:diagrams_delta}}
\end{figure}

Contributions of the $\Delta$ intermediate state to the tree-level charged-current elastic scattering amplitudes on nucleons $T^\Delta_{\nu_\ell n \to \ell^- p}$ and $T^\Delta_{\overline{\nu}_\ell p \to \ell^+ n}$ are given by
\begin{align}
	T^\Delta_{\nu_\ell n \to \ell^- p} &= e^2 \sqrt{2} G_F V_{ud} \int \hspace{-0.1cm} \frac{\mathrm{d}^d L}{ \left( 2 \pi \right)^d} \overline{\ell} \gamma^\mu \frac{ - \slash{p}' - \slash{L} - m_\ell}{\left(L+p'\right)^2-m_\ell^2} \gamma_\sigma \mathrm{P}_\mathrm{L} \nu_{\ell} \mathrm{\Pi}_{\mu \nu} \left( L \right) \nonumber \\
& \overline{p} \left [ \left( \gamma^0 J^{\nu \beta}_{V, N \to \Delta} \gamma^0 \right)^\dagger \frac{\slash{k}' - \slash{L} + M_\Delta}{\left( L - k'\right)^2 - M_\Delta^2} \mathrm{\Pi}^\Delta_{\beta \alpha} J^{\sigma \alpha}_{N \to \Delta} + \left( \gamma^0 J^{\sigma \beta}_{N \to \Delta} \gamma^0 \right)^\dagger \frac{\slash{k} + \slash{L} + M_\Delta}{\left( L + k\right)^2 - M_\Delta^2} \mathrm{\Pi}^\Delta_{\beta \alpha} J^{\nu \alpha}_{V, N \to \Delta} \right] n, \label{eq:contribution_of_boxes1} \\
	T^\Delta_{\overline{\nu}_\ell p \to \ell^+ n} &= e^2 \sqrt{2} G_F V_{ud} \int \hspace{-0.1cm} \frac{\mathrm{d}^d L}{ \left( 2 \pi \right)^d} \overline{\overline{\nu}}_\ell \gamma_\sigma \mathrm{P}_\mathrm{L} \frac{ \slash{p}'+ \slash{L} + m_\ell}{\left(L+p'\right)^2-m_\ell^2} \gamma^\mu \overline{\ell} \mathrm{\Pi}_{\mu \nu} \left( L \right) \nonumber \\
& \overline{n} \left [ \left( \gamma^0 J^{\nu \beta}_{V, N \to \Delta} \gamma^0 \right)^\dagger \frac{\slash{k}' - \slash{L} + M_\Delta}{\left( L - k'\right)^2 - M_\Delta^2} \mathrm{\Pi}^\Delta_{\beta \alpha} J^{\sigma \alpha}_{N \to \Delta} + \left( \gamma^0 J^{\sigma \beta}_{N \to \Delta} \gamma^0 \right)^\dagger \frac{\slash{k} + \slash{L} + M_\Delta}{\left( L + k\right)^2 - M_\Delta^2} \mathrm{\Pi}^\Delta_{\beta \alpha} J^{\nu \alpha}_{V, N \to \Delta} \right] p, \label{eq:contribution_of_boxes2}
\end{align}
with the $\Delta$ mass $M_\Delta$, the projection operator on the spin-$3/2$ states in the $\Delta$ propagator $\mathrm{\Pi}^\Delta_{\beta \alpha}$:
\begin{equation} \label{eq:delta_projection_propagator}
	\mathrm{\Pi}^\Delta_{\beta \alpha} = - g_{\beta \alpha} + \frac{1}{3} \gamma_\beta \gamma_\alpha + \frac{\slash{p}_\Delta \gamma_\beta \left( p_\Delta \right)_\alpha + \left( p_\Delta \right)_\beta \gamma_\alpha \slash{p}_\Delta}{3 p_\Delta^2},
\end{equation}
and the momentum-space photon propagator $\mathrm{\Pi}^{\mu \nu}$:
\begin{equation} \label{eq:photon_propagator}
	\mathrm{\Pi}^{\mu \nu} \left( L \right) = \frac{i}{ L^2 - \lambda_\gamma^2}\left( - g^{\mu \nu} + \left( 1 - \xi_\gamma \right) \frac{L^\mu L^\nu}{L^2 - a \xi_\gamma \lambda_\gamma^2}\right), 
\end{equation}
with the photon-mass regulator $\lambda_\gamma$, the gauge-fixing parameter $\xi_\gamma$, and an arbitrary constant $a$. The transition currents are described in Eqs.~(\ref{eq:form_factors_decomposition1})-(\ref{eq:vector_transition_form_factors}). We illustrate the corresponding diagrams in Fig.~\ref{fig:diagrams_delta}.

In contrast to the QED radiative corrections in Refs.~\cite{Tomalak:2021hec,Tomalak:2022xup} with nucleon intermediate states only, the resonance model with magnetic dipole transition of Eq.~(\ref{eq:vector_transition_form_factors}) is gauge invariant, does not depend on the arbitrary regularization parameter $a$, and is free from soft and collinear divergences. The product of two vector transitions does not depend on the momentum-dependent part of the $\Delta$ propagator $\mathrm{\Pi}^\Delta_{\beta \alpha}$, while the axial-vector transition current $J^{\sigma \alpha}_{A, N \to \Delta}$ is contracted with $\left( p_\Delta \right)_\alpha$ that results in $ p_\Delta^{\sigma} C_5^A + \frac{ p_\Delta \cdot q }{M} \frac{q^{\sigma}}{M} C_6^A $. For this contribution, the resonance propagator $\frac{\slash{p}_\Delta+ M_\Delta}{p^2_\Delta - M^2_\Delta}$ is modified to $\frac{1}{p^2_\Delta} + \frac{\slash{p}_\Delta+ M_\Delta}{M_\Delta} \left( \frac{1}{p^2_\Delta - M^2_\Delta} - \frac{1}{p^2_\Delta} \right) = \frac{1}{M_\Delta} \frac{\slash{p}_\Delta+ M_\Delta}{p^2_\Delta - M^2_\Delta} - \frac{1}{M_\Delta} \frac{\slash{p}_\Delta+ M_\Delta}{p^2_\Delta } + \frac{1}{p^2_\Delta }$, which we implement in our calculation. In the hadronic model below, we reduce the number of loop momenta in the contribution from the form factor $C_6^A$ before performing the integration by using $p_\Delta \cdot q = \frac{p_\Delta^2 - M^2 + q^2}{2}$ as 
\begin{align}
	& \frac{ p_\Delta \cdot q }{M} \frac{q^{\sigma}}{M} \left( \frac{1}{M_\Delta} \frac{\slash{p}_\Delta+ M_\Delta}{p^2_\Delta - M^2_\Delta} - \frac{1}{M_\Delta} \frac{\slash{p}_\Delta+ M_\Delta}{p^2_\Delta } + \frac{1}{p^2_\Delta } \right) = \nonumber \\
& \frac{ q^2 }{2 M^2} \frac{q^{\sigma}}{M_\Delta} \frac{\slash{p}_\Delta+ M_\Delta}{p^2_\Delta - M^2_\Delta} + \frac{ q^2 - M^2}{2 M^2} \frac{q^{\sigma}}{M_\Delta} \left( - \frac{\slash{p}_\Delta}{p^2_\Delta } \right) + \frac{1}{2 M} \frac{q^{\sigma}}{M}.
\end{align}

To investigate the dependence on hadronic physics modeling, we perform two calculations with the argument of form factors in the electroweak transition vertex either $(q \pm L)^2$, the hadronic model $\Delta^{\mathrm{hm}}_{\mathrm{I},\mathrm{II}}$, or $q^2$, the default model $\Delta^0_{\mathrm{I},\mathrm{II}}$. We calculate the $\Delta$-resonance contribution to all invariant amplitudes in Eqs.~(\ref{eq:CCQE_amplitude1})-(\ref{eq:CCQE_amplitudem2}) by evaluating the ultraviolet-finite helicity amplitudes of Eqs.~(\ref{eq:contribution_of_boxes1}) and~(\ref{eq:contribution_of_boxes2}) in $d=4$. We verify the imaginary parts of the invariant amplitudes for incoming neutrino energies $E_\nu \ge \frac{\left(M_\Delta + m_\ell\right)^2 - M^2}{2M}$ through an independent unitarity calculation. As expected, the collinear contribution to the amplitudes $g_M,~g_E,~f_A,$ and $\fAt$ ($f_P,~f_{3},~f_R,$ and $f_T$) vanishes (scales as $\ln m_\ell$) in the limit of vanishing lepton mass, i.e., $m_\ell \to 0$, respectively.

The one-loop contribution from the product of two vector transitions, as well as from the momentum-dependent piece of the $\Delta$ propagator $\mathrm{\Pi}^\Delta_{\beta \alpha}$ in the hadronic model, exhibits an additional crossing symmetry. The crossed contributions to invariant amplitudes $f^\mathrm{x}_i,~g^\mathrm{x}_i$, the second terms in expressions between nucleon spinors in Eqs.~(\ref{eq:contribution_of_boxes1}) and~(\ref{eq:contribution_of_boxes2}), can be obtained from the direct contributions to invariant amplitudes $f^\mathrm{d}_i,~g^\mathrm{d}_i$, the first terms in expressions between nucleon spinors in Eqs.~(\ref{eq:contribution_of_boxes1}) and~(\ref{eq:contribution_of_boxes2}), by applying the following crossing relations
\begin{align}
	\mathfrak{Re} \left[ g^\mathrm{x}_M, g^\mathrm{x}_E \right] \left( \nu,~t \right) &= - \mathfrak{Re} \left[ g^\mathrm{d}_M, g^\mathrm{d}_E \right]\left( -\nu,~t \right), \\
	\mathfrak{Re} \left[ f^\mathrm{x}_2, f^\mathrm{x}_{A3}, f^\mathrm{x}_{T} \right]\left( \nu,~t \right) &= - \mathfrak{Re} \left[ f^\mathrm{d}_{2}, f^\mathrm{d}_{A3}, f^\mathrm{d}_T \right]\left( -\nu,~t \right), \\
	\mathfrak{Re} \left[ f^\mathrm{x}_A, f^\mathrm{x}_P, f^\mathrm{x}_{R}, f^\mathrm{x}_{3} \right]\left( \nu,~t \right) &= \mathfrak{Re} \left[ f^\mathrm{d}_{A}, f^\mathrm{d}_{P}, f^\mathrm{d}_{R}, f^\mathrm{d}_3 \right]\left( -\nu,~t \right).
\end{align}

\begin{figure}[t]
\centering
\includegraphics[height=0.674\textwidth]{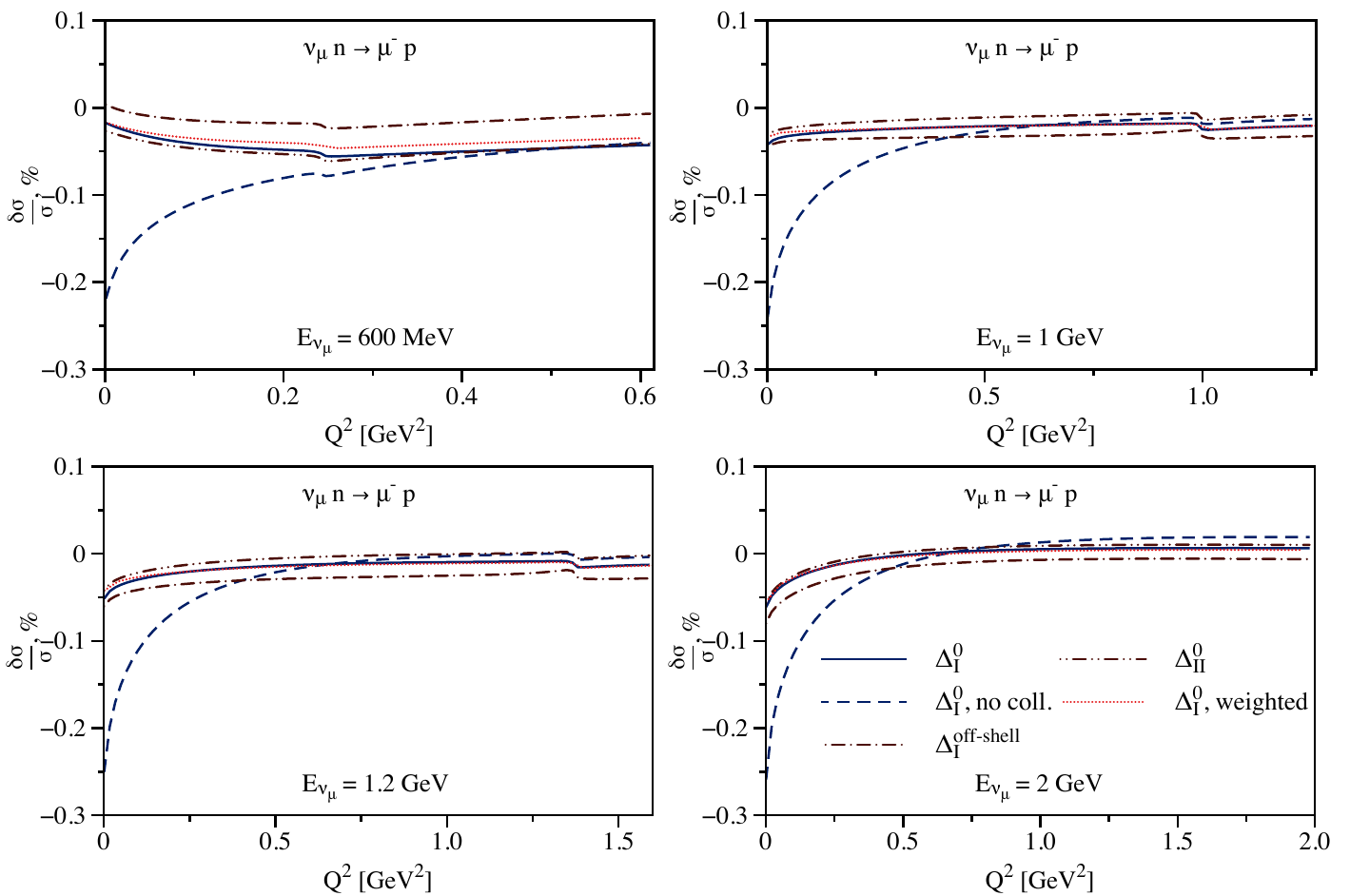}
\caption{Relative contribution of the $\Delta$ intermediate state to the unpolarized charged-current muon neutrino-neutron elastic scattering cross section is shown as a function of the squared momentum transfer $Q^2$ for initial neutrino energies $E_\nu = 600~\mathrm{MeV}$ (upper left plot), $E_\nu = 1~\mathrm{GeV}$ (upper right plot), $E_\nu = 1.2~\mathrm{GeV}$ (lower left plot), and $E_\nu = 2~\mathrm{GeV}$ (lower right plot). Results for different model calculations, which are described in Section~\ref{sec:virtual_delta}, are presented. \label{fig:muon_neutrino_virtual_delta}}
\end{figure}

\begin{figure}[t]
\centering
\includegraphics[height=0.674\textwidth]{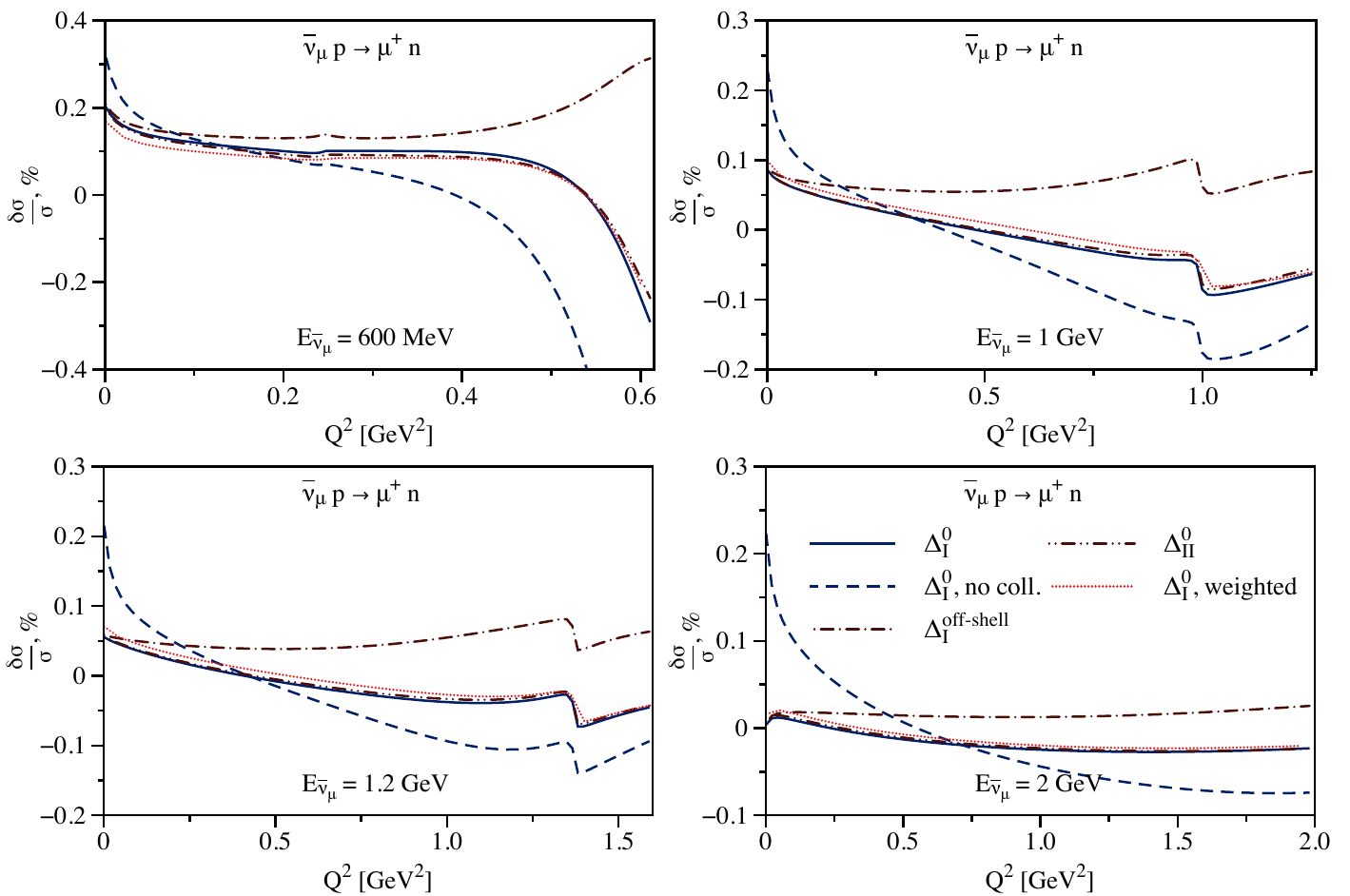}
\caption{Same as Figure~\ref{fig:muon_neutrino_virtual_delta} but for the antineutrino-proton scattering. \label{fig:muon_antineutrino_virtual_delta}}
\end{figure}

\begin{figure}[t]
\centering
\includegraphics[height=0.674\textwidth]{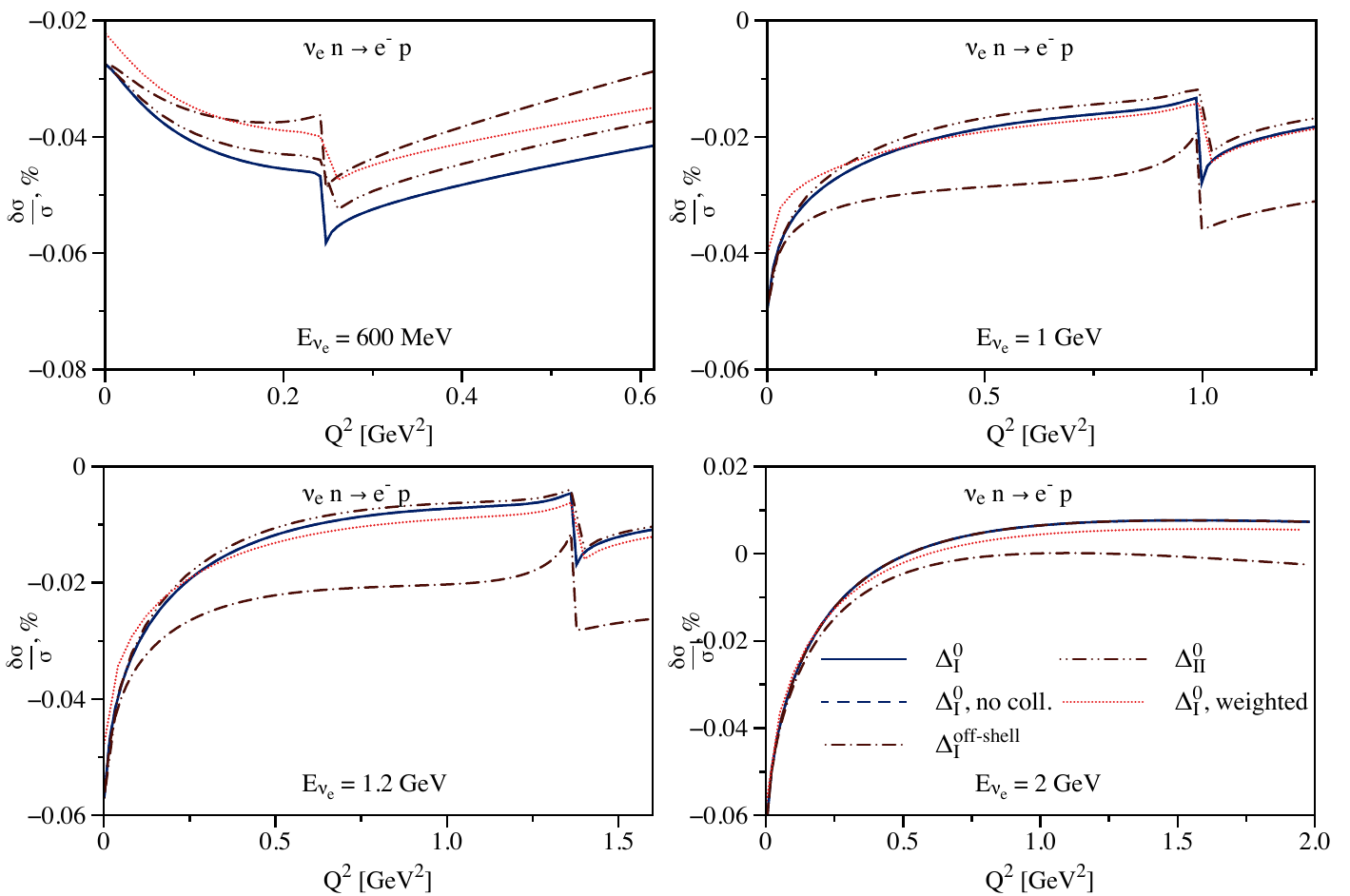}
\caption{Same as Figure~\ref{fig:muon_neutrino_virtual_delta} but for the electron flavor. \label{fig:electron_neutrino_virtual_delta}}
\end{figure}

\begin{figure}[t]
\centering
\includegraphics[height=0.674\textwidth]{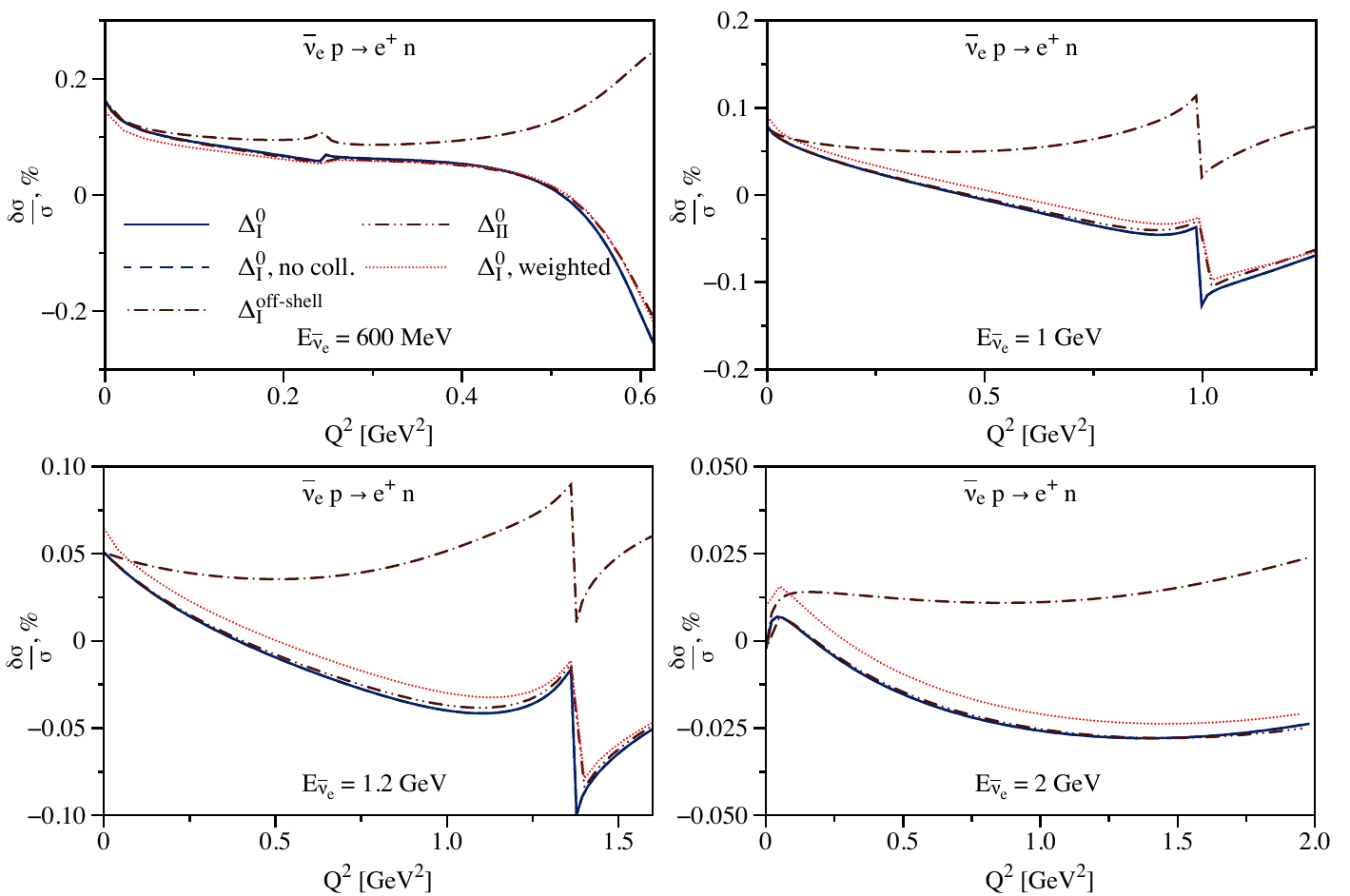}
\caption{Same as Figure~\ref{fig:muon_antineutrino_virtual_delta} but for the electron flavor. \label{fig:electron_antineutrino_virtual_delta}}
\end{figure}

The leading $\Delta$-resonance contribution to the unpolarized cross section arises from the interference of radiative corrections to invariant amplitudes with the tree-level form factors in Eq.~(\ref{eq:xsection_CCQE}). For the nucleon form factors, we take default fits of the experimental data with $Q^2 < 1~\mathrm{GeV}^2$ from the analysis of electron and neutrino scattering as well as atomic spectroscopy~\cite{Meyer:2016oeg,Borah:2020gte}. Evaluating the loop contributions, we specify the $N\to \Delta$ transition form factors as described in Section~\ref{sec:delta_production}. We evaluate the relative $\Delta$-resonance contribution to the tree-level cross sections for (anti)neutrino scattering on a nucleon target, considering incoming (anti)neutrino energies $E_\nu = 600~\mathrm{MeV},~1~\mathrm{GeV},~1.2~\mathrm{GeV},$ and $2~\mathrm{GeV}$. We present our results in Figures~\ref{fig:muon_neutrino_virtual_delta} and~\ref{fig:muon_antineutrino_virtual_delta} for muon flavor and in Figures~\ref{fig:electron_neutrino_virtual_delta} and~\ref{fig:electron_antineutrino_virtual_delta} for electron flavor, respectively.

To investigate uncertainties arising from the unknown off-shell $N \to \Delta$ transition form factors, we perform an additional calculation for the default model, but using the decomposition of the vector part for the $N\to \Delta$ transition from Eq.~(\ref{eq:form_factors_decomposition1}) with the form factors from Eqs.~(\ref{eq:three_vector_transition_form_factors1})-(\ref{eq:three_vector_transition_form_factors3}) and $\xi_\gamma = 1$ for the contribution from the transition form factor $C_5^V$ and call this calculation $\Delta^\mathrm{off-shell}_{\mathrm{I}}$. We compare $\Delta^\mathrm{off-shell}_{\mathrm{I}}$ results to the calculations $\Delta^0_{\mathrm{I},\mathrm{II}}$ in Figures~\ref{fig:muon_neutrino_virtual_delta}-\ref{fig:electron_antineutrino_virtual_delta}.

The results for $\Delta^0_\mathrm{I}$ and $\Delta^0_\mathrm{II}$, calculated using different axial-vector transition form factors, cf. Eq.~(\ref{eq:delta_model}) and the description above, are in very close agreement. This indicates the dominant role of the electroweak axial-vector contribution from the form factor $C_5^A$. The subtraction of power-suppressed collinear contributions has a negligible impact on the results, which validates the approximate flavor independence for the resonance contributions to the unpolarized cross section. For the electron flavor, this subtraction is negligible compared to the magnitude of the $\Delta$-resonance contribution. However, it is more significant for the muon flavor due to power corrections that involve the muon mass. To explicitly study the mass dependence, we subtract the power-suppressed collinear region from each loop integral in the default model while keeping the kinematic relations between the amplitudes and loop integrals exact. We refer to this calculation as ``$\Delta^0_\mathrm{I},~\mathrm{no~coll.}$", and compare it with the results obtained without any subtraction.

Nevertheless, the $\Delta (1232)$-resonance contribution demonstrates an expected kinematic enhancement, as shown in Figures~\ref{fig:muon_neutrino_virtual_delta}-\ref{fig:electron_antineutrino_virtual_delta}, the resulting correction does not exceed the permille level. For the dominant region of (anti)neutrino fluxes in oscillation experiments, the kinematic enhancement happens away from the near-forward region, $Q^2 \approx 0$, with the largest cross sections. For further validation, we also perform a calculation in the hadronic model $\Delta^{\mathrm{hm}}_{\mathrm{I},\mathrm{II}}$ and find a permille-level correction with a larger kinematic enhancement in a relatively limited phase-space region. The differences between model calculations for the $\Delta$-resonance contribution to the unpolarized elastic (anti)neutrino-nucleon scattering are consistent with the uncertainty estimates in Refs.~\cite{Tomalak:2021hec,Tomalak:2022xup}.

We also perform a weighted-$\Delta$ calculation by integrating the cross-section correction over the invariant mass of the $\Delta$ resonance
\begin{equation}
	\sigma^\Delta \left( M_\Delta \right) \to \int \limits_{M + m_\pi}^{\sqrt{s}} f \left( W \right) \sigma^\Delta \left( W \right) \mathrm{d} W,
\end{equation}
where the weighting function $f(W)$ is given by the relativistic Breit-Wigner distribution~\cite{Tomalak:2017shs}:
\begin{equation} \label{eq:narrow_delta_weight}
	f \left( W \right) = \frac{N_\Delta}{W^6} \frac{\left( W^2 - M^2 + m^2_\pi \right)^2 - 4 W^2 m^2_\pi}{\left( W^2 - M_\Delta^2 \right)^2 + M^2_\Delta \Gamma_\Delta^2} \Theta \left( W - M - m_\pi \right),
\end{equation}
with the pion mass $ m_\pi \approx 0.135~\mathrm{GeV}$, the $\Delta$ mass $ M_\Delta = 1.232~\mathrm{GeV}$, and the $\Delta$ width $\Gamma_\Delta = 0.117~\mathrm{GeV}$. The normalization parameter $N_\Delta$ is chosen to satisfy $\int \limits_{M + m_\pi}^{\infty} f \left(W\right) \mathrm{d} W = 1$. We present the weighted-$\Delta$ calculation in Figures~\ref{fig:muon_neutrino_virtual_delta}-\ref{fig:electron_antineutrino_virtual_delta} and find that the narrow-$\Delta$ approximation agrees well with the weighted-$\Delta$ result.

Our study without enforcing the consistent couplings~\cite{Pascalutsa:1998pw,Pascalutsa:2000kd,Pascalutsa:2006up,Pascalutsa:2007wz} finds a $1$-$3$ permille-level dependence on the treatment of off-shell contributions in the radiative $\Delta$ decay, which rules out predictions based on the gauge-dependent decomposition of vector contributions in Eq.~(\ref{eq:form_factors_decomposition1}) more precise than order-of-magnitude estimates for both moderate and large momentum transfers. This result is in qualitative agreement with findings for the radiative $\Delta$ decay in neutral-current process~\cite{Hill:2009ek,Hill:2010zy}. We therefore provide an independent uncertainty estimate for background channels in searches for new physics at short-baseline neutrino experiments~\cite{LSND:1996ubh,LSND:1997vun,Blanpied:1997zz,MiniBooNE:2007uho,MiniBooNE:2008yuf,Sparveris:2008jx,MiniBooNE:2009atu,MiniBooNE:2013uba,Wang:2013wva,Garvey:2014exa,Wang:2014nat,Wang:2015ivq,Rosner:2015fwa,MiniBooNE:2018esg,Machado:2019oxb,T2K:2019odo,Ioannisian:2019kse,Giunti:2019sag,MiniBooNE:2020pnu,Chanfray:2021wie,MicroBooNE:2021zai,Goncalves:2022lmi,MicroBooNE:2025rsd,MicroBooNE:2025ovj,Ioannisian:2025bro}.

\section{Resonance contributions to radiative corrections in factorization framework}
\label{sec:factorization_framework}

This Section presents the relationship between the present calculation and the factorization approach to QED radiative corrections for charged-current elastic (anti)neutrino-nucleon scattering developed in Refs.~\cite{Tomalak:2021hec,Tomalak:2022xup}.

In this framework, applicable at accelerator (anti)neutrino energies, the differential cross section $\mathrm{d} \sigma$ factorizes into hard ($H$), jet ($J$), and soft ($S$) functions as
\begin{equation} \label{eq:factorization_formula}
	\mathrm{d} \sigma \sim H \left( \frac{\mu}{\Lambda} \right) J \left( \frac{\mu}{m_\ell} \right) S \left( \frac{\mu}{\Delta E} \right),
\end{equation}
where $\mu$ is the common renormalization scale, $\Lambda \sim M_\Delta \sim M \sim E_\nu \sim \sqrt{Q^2}$ is the hard scale, and $\Delta E$ is the soft-photon energy acceptance parameter, which defines the maximum energy of photons treated as soft radiation. The soft and jet functions are perturbative and encapsulate the universal long-distance physics responsible for large logarithmic enhancements, which correspond to infrared and collinear singularities. These logarithms become significant at the hard scale $\mu \sim \Lambda$. In contrast, the hard function is nonperturbative and describes QED interactions of hadrons.

At leading order, the hard function $H^\mathrm{0}$ is expressed in terms of nucleon form factors. The dominant QED correction $H^\mathrm{N}$ from the nucleon intermediate states was computed in Refs.~\cite{Tomalak:2021hec,Tomalak:2022xup}. The primary goal of this work is to calculate the contribution of the $\Delta(1232)$ resonance, $H^\Delta$, thereby extending the hard function as
\begin{equation} \label{eq:extended_hard_function}
	H = H^\mathrm{0} + H^\mathrm{N} + H^\mathrm{\Delta}.
\end{equation}

Our analysis demonstrates that the magnitude of $H^\mathrm{\Delta}$ is comparable to the dominant theoretical uncertainty associated with inelastic intermediate states in the previously calculated $H^\mathrm{0} + H^\mathrm{N}$ in Refs.~\cite{Tomalak:2021hec,Tomalak:2022xup}. The logarithmic enhancement from $H^\mathrm{\Delta}$ enters the scattering cross section as power-suppressed contributions, resulting in a permille-level virtual correction. An analogous suppression pattern also applies for the $\Delta$-induced radiation of real photons, as these contributions are free from soft and collinear singularities at $\mathcal{O} \left(e^2 \right)$. Consequently, while the $\Delta$ resonance constitutes a non-negligible component of the hard function, its impact on the overall QED radiative corrections remains sub-dominant compared to the leading logarithmic enhancements captured by the functions $S$ and $J$.

\section{Conclusions and outlook}
\label{sec:conclusions}

In this study, we evaluated the $\Delta(1232)$-resonance contribution to the charged-current elastic (anti)neutrino-nucleon scattering for the first time at accelerator neutrino energies. We modeled the $\Delta$ interaction vertices in the on-shell form, with and without shifts in momenta for the electroweak interaction vertex. We approximated the vector part of the nucleon-to-delta transition current by the leading magnetic dipole transition. We determined the virtual contributions of the $\Delta$ intermediate state to all invariant amplitudes. Throughout the calculation, we verified that the resulting amplitudes respected the expected soft, collinear, and crossing symmetry properties. We also ensured the consistency with unitarity conditions for the invariant amplitudes of inelastic excitations in the charged-current elastic (anti)neutrino-nucleon scattering. Our results show that the $\Delta$-resonance contribution remains at the permille level for the unpolarized elastic scattering cross sections, which is in agreement with the error estimates presented in previous works~\cite{Tomalak:2021hec,Tomalak:2022xup}. The analysis further confirms no significant kinematic enhancements of the resonance contribution. Not only do the central values of radiative corrections are validated, but the corresponding uncertainties from the earlier studies are also confirmed, ensuring that the predictions made in Refs.~\cite{Tomalak:2021hec,Tomalak:2022xup} remain the most precise and reliable for modern and future experiments and studies.

\appendix

\section{Relation between helicity and invariant amplitudes}
\label{app:helicity_and_invariants}

To describe charged-current elastic (anti)neutrino-nucleon scattering, there are $8$ helicity amplitudes $ T_{\lambda^\prime h^\prime, \lambda} $ with arbitrary helicities of the charged lepton, final and initial nucleons $h^\prime, \lambda, \lambda^\prime$ = $\pm 1/2$, respectively: $ T_1 = T_{ \hp \hp, \hp}, ~T_2 = T_{\hp \hm, \hp}, ~T_3 = T_{\hm \hp, \hm}, ~T_4 = T_{\hm \hm, \hm}, ~T_5 = T_{\hp \hp, \hm}, ~T_6 = T_{\hp \hm, \hm}, ~T_7 = T_{\hm \hp, \hp}, ~T_8 = T_{\hm \hm, \hp} $.

Using the Jacob and Wick~\cite{Jacob:1959at} phase convention for the spinors, the helicity amplitudes $ T_{\lambda^\prime h^\prime, \lambda} $ for elastic (anti)neutrino-nucleon scattering can be expressed in terms of the invariant amplitudes as
\begin{align}
	T_1 \left( x_1, x_2, \theta \right) &= \left( \omega f_R + 4 \left( g_M + f_A - f_3 \right) + 2 \left( f_T - 2 \fAt \right)\right) r_\ell x_1 \sin \theta \nonumber \\
&+ \left( 2 \left( f_1 + f_A \right) - 2 \omega \left( f_2 - 2 \fAt \right) + r_\ell^2 \left( 2 \left( f_2 - 2 \fAt \right) - f_R - 4 \left( f_P + f_3 \right) \right) \right) x_2 \sin \theta, \\
	T_2 \left( x_1, x_2, \theta \right) &= \left( \left( 1 + 2 \omega \right) \left( 2 \left( f_1 + f_A \right) + r_\ell^2 f_R \right) + r_\ell^2 \left( 2 \left( f_2 + 2 \fAt \right) + 4 \left( f_P - f_3 + f_T\right) \right) \right) x_1 \frac{1 + \cos \theta}{\sqrt{1 + 2 \omega}} \nonumber \\
&- \left( \omega \left( 2 \left(f_2 - 2 \fAt \right) +4 \left( f_P + f_3 \right) \right) + 4 f_3 - 2 \left( 2 \fAt + f_T \right) \right) r_\ell x_2 \frac{1 + \cos \theta}{\sqrt{1 + 2 \omega}}, \\
	T_3 \left( x_1, x_2, \theta \right) &= - \left( \omega f_R + 4 f_3 - 2 \left( 2 \fAt + f_T \right) \right) r_\ell x_1 \sin \theta \nonumber \\
&+ \left( 2 \left( f_1 + f_A \right) - 2 \omega \left( f_2 + 2 \fAt \right) + r_\ell^2 \left( 2 \left( f_2 + 2 \fAt \right) + f_R + 4 \left( f_P - f_3 + f_T \right) \right) \right) x_2 \sin \theta, \\
	T_4 \left( x_1, x_2, \theta \right) &= \left( \left( 1 + 2 \omega \right) \left( 2 \left( f_1 - f_A \right) - r_\ell^2 f_R \right) + r_\ell^2 \left( 2 \left( f_2 - 2 \fAt \right) - 4 \left( f_P + f_3 \right) \right) \right) x_1 \frac{1 + \cos \theta}{\sqrt{1 + 2 \omega}} \nonumber \\
&- \left( \omega \left(2 \left( f_2 + 2 \fAt \right) - 4 \left( f_P - f_3 \right) \right) +4 \fAt \right) r_\ell x_2 \frac{1 + \cos \theta}{\sqrt{1 + 2 \omega}} \nonumber \\
&+ 4 \left( \omega \left( g_M - f_A \right) - f_A \right) x_1 \frac{1 - \cos \theta}{\sqrt{1 + 2 \omega}} \nonumber \\
&- \left( 4 \left( g_M + f_A - f_3 \right) + 2 \left( 1 + 2 \omega \right) f_T \right) r_\ell x_2 \frac{1 - \cos \theta}{\sqrt{1 + 2 \omega}} - 4 \left( \left( 1 + 2 \omega \right) f_T + 2 f_3 \right) r_\ell x_2, \\
	T_5 \left( x_1, x_2, \theta \right) &= - T_4 \left( x_2, x_1, \pi - \theta \right), \\
	T_6 \left( x_1, x_2, \theta \right) &= - T_3 \left( x_2, x_1, \pi - \theta \right), \\
	T_7 \left( x_1, x_2, \theta \right) &= T_2 \left( x_2, x_1, \pi - \theta \right), \\
	T_8 \left( x_1, x_2, \theta \right) &= T_1 \left( x_2, x_1, \pi - \theta \right),
\end{align}
with the scattering angle in the center-of-mass reference frame $\theta$ and kinematic notations:
\begin{align}
	x_1 + x_2 &= \frac{M \sqrt{ \left( s - M^2 \right) \left( s - \left( M - m \right)^2 \right) }}{2\sqrt{s}}, \\
	x_1 - x_2 &= \frac{M \sqrt{ \left( s - M^2 \right) \left( s - \left( M + m \right)^2 \right) }}{2\sqrt{s}}, \\
	\omega &= \frac{s-M^2}{2 M^2}.
\end{align}

The same relations hold for the antineutrino scattering after redefining $ T_1 = T_{ \hm \hm, \hm}, ~T_2 = -T_{\hm \hp, \hm}, ~T_3 = T_{\hp \hm, \hp} , ~T_4 = -T_{\hp \hp, \hp}, ~T_5 = - T_{\hm \hm, \hp}, ~T_6 = T_{\hm \hp, \hp}, ~T_7 = -T_{\hp \hm, \hm}, ~T_8 = T_{\hp \hp, \hm}$, equivalently $T_{\lambda^\prime h^\prime, \lambda} \to \left( - 1 \right)^{8 \lambda \lambda^\prime h^\prime} T_{-\lambda^\prime -h^\prime, -\lambda}$, and changing the sign for the amplitudes $f_A$ and $f_{3}$.

These relations can be inverted to express the invariant amplitudes in terms of the helicity amplitudes. We present the inverted relations in the Supplementary Material.

\section*{Acknowledgements}

Author thanks Barbara Pasquini, Marc Vanderhaeghen, and Lothar Tiator for numerous discussions while working with pion electroproduction amplitudes, Kevin Quirion and Emilie Passemar for numerous discussions while working with pion neutrinoproduction. This work is supported by the US Department of Energy through the Los Alamos National Laboratory and by LANL’s Laboratory Directed Research and Development (LDRD/PRD) program under project numbers 20210968PRD4 and 20240127ER. Los Alamos National Laboratory is operated by Triad National Security, LLC, for the National Nuclear Security Administration of the US Department of Energy (Contract No. 89233218CNA000001). FeynCalc~\cite{Mertig:1990an,Shtabovenko:2016sxi}, LoopTools~\cite{Hahn:1998yk}, Mathematica~\cite{Mathematica}, and DataGraph~\cite{JSSv047s02} were extremely useful in this work. This work is supported by the National Science Foundation of China under Grants No. 12347105, No. 12447101.

\newpage

\bibliography{resonances}{}

\end{document}